\documentclass[aip,apl,reprint,twocolumn]{revtex4-1}
\usepackage{graphicx}% Include figure files
\usepackage{dcolumn}% Align table columns on decimal point
\usepackage{bm}% bold math
\usepackage{amssymb,amsmath}
\usepackage{gensymb}
\usepackage{balance}
\usepackage{textcomp}
\usepackage{times,mathptmx}
\usepackage{graphicx}
\usepackage{lastpage}
\bibstyle{apsrev}
\usepackage{hyperref}
\usepackage{color}
\usepackage[normalem]{ulem}
\begin{document}
%\draft
\preprint{PREPRINT}

\title[Molecular Dynamics Investigation of the Influence of the Shape of Cation on the Structure and Lubrication Properties of Ionic Liquids]{Molecular Dynamics Investigation of the Influence of the Shape of Cation on the Structure and Lubrication Properties of Ionic Liquids}

%Phase diagram of dipolar tubes: ferromagnetic and antiferromagnetic states .. Phase diagram and cohesive energy of dipolar tubes: ferromagnetic and antiferromagnetic states

%1st author
\author{Miljan Da\v{s}i\'{c}}
\affiliation{
Scientific Computing Laboratory, Center for the Study of Complex Systems, Institute of Physics Belgrade, University of Belgrade, 11080 Belgrade, Serbia.
}
%2nd Author
\author{Igor Stankovi\'{c}}
\affiliation{
Scientific Computing Laboratory, Center for the Study of Complex Systems, Institute of Physics Belgrade, University of Belgrade, 11080 Belgrade, Serbia.
}
\email{igor.stankovic@ipb.ac.rs}

%3th author
\author{Konstantinos Gkagkas}
\affiliation{Advanced Technology Division, Toyota Motor Europe NV/SA, Technical Center, Hoge Wei 33B, 1930 Zaventem, Belgium.}

\date{\today}% It is always \today, today, but you may specify any date with
             % \date.

%%%%%%%%%%%%%%%%%%%%%%%%%%%%%%%%%%%%%%%%%%%%%%%%%%%%%%%%%%%%%
\begin{abstract}
We present a theoretical study of the influence of the molecular geometry of the cation on the response of ionic liquid (IL) to confinement and mechanical strain. The so-called {\it tailed} model includes a large spherical anion and asymmetric cation consisting of a charged head and neutral tail. Despite its simplicity, this model recovers a wide range of structures seen in IL: simple cubic lattice for the small tails, liquid-like state for symmetric cation-tail dimers, and molecular layer structure for dimers with large tails. A common feature of all investigated model ILs is the formation of a fixed (stable) layer of cations along solid plates. We observe a single anionic layer for small gap widths, a double anionic layer for intermediate ones, and tail--to--tail layer formation for wide gaps. The normal force evolution with the gap size can be related to the layer formed inside the gap. The low hysteretic losses during the linear cyclic motion suggest the presence of strong slip inside the gap. In our model the specific friction is low and friction force decreases with the tail size.\end{abstract}

%%%END OF FOOTNOTES%%%
\maketitle
 
\vspace{0.6cm}
\section{Introduction}
Ionic liquids (ILs) are two-component systems composed of large asymmetric and irregularly shaped organic cations and anions. 
The feature of irregularity is important as it is effectively preventing low-temperature ordering and crystallisation. 
Therefore, ILs are usually in the melted or glassy state. 
Physical properties of ILs like negligible vapour pressure, high-temperature stability, high ionic conductivity and also a great variety of ILs, and their mixtures highlight them as potentially relevant to lubrication~\cite{zhou2009ionic,hayes2010interface}.
A large number of variations in IL composition is possible, estimated at the order of magnitude of $10^{18}$ different ILs~\cite{dold2013influence}. 
From their variety stems the possibility of tuning their physicochemical properties which can affect lubrication such as viscosity, polarity, surface reactivity by varying their atomic composition, 
as well as the cation-anion combination. Hence, it would be advantageous if we could deduce general relations between the molecular structure and anti-wear and lubrication properties of ILs.

Since 2001, when ionic liquids were first considered for lubrication applications~\cite{liu2002tribological}, there has been a large number of experimental studies in that direction. 
It has been observed that the alkyl chain length of the cations affects the IL viscosity~\cite{zhou2009ionic}, melting point~\cite{zhou2009ionic} and pressure-viscosity coefficients~\cite{pensado2008pressure}. 
Related specifically to lubrication, \citeauthor{dold2013influence}~\cite{dold2013influence} and \citeauthor{minami2009ionic}~\cite{minami2009ionic} explored the impact of cationic alkyl chain's length
on the tribological properties of ILs. ILs considered in those references have the same cations but different anions: symmetric hexafluorophosphate $\left[PF_6\right]^{-}$ and asymmetric bis(trifluoromethylsulfonyl) imide $\left[Tf_2N\right]^{-}$, respectively.
Still, while \citeauthor{minami2009ionic} observed that the coefficient of friction (COF) decreases from $0.25$ to $0.15$ with the increase of alkyl chain length $n_{\rm C} = 2$ to $12$ ($n_{\rm C}$ is the number of carbon atoms), 
\citeauthor{dold2013influence} observed that the COF increases from $0.025$ to $0.1$. The IL's wetting properties are also sensitive to its molecular geometry. ILs change wetting behaviour depending on the anion size~\cite{bou2010nanoconfined,beattie2013molecularly,wang2013impact}: from the absence of wetting to partial or complete wetting.  A well--studied IL $\left[BMIM\right]^{+}\left[PF_6\right]^{-}$ exhibits full wetting at the interface with mica substrates~\cite{bou2010nanoconfined,beattie2013molecularly}. 
On contrary, $\left[BMIM\right]^{+}\left[TFSI\right]^{-}$ shows partial wetting on mica~\cite{wang2013impact,beattie2013molecularly}. 
In these examples, ILs have the same cation and different anions.
 
An important observation about the structure of confined ILs is their arrangement into positively and negatively charged ionic layers and adsorption onto solid surfaces~\cite{smith2013quantized, lubricants2013}.
These ionic adsorption layers should reduce friction and prevent wear, especially in the case of boundary lubrication~\cite{smith2013quantized}. 
The wear is reduced primarily in two ways: via high load-carrying capability and self-healing of adsorbed IL layers. 
Still, these two processes seem conflicting with each other since high load-carrying capability requires strong adsorption of the lubricant to the surface while self-healing requires high mobility~\cite{bhushan1995nanotribology}.
Understanding the driving forces between them requires relating the molecular structure and flow properties of confined IL.  
\citeauthor{kamimura2006relationship}~\cite{kamimura2006relationship} have evaluated tribological properties of different ionic liquids by pendulum and ball on disk tribo testers.
They have considered ILs consisting of imidazolium cations with different alkyl chain length and $\left[Tf_2N\right]^{-}$ anion as lubricants. 
Their main observation is that the increment of alkyl chain length can reduce friction and wear of sliding pairs in the elastohydrodynamic lubrication regime (EHL) as a consequence of increased viscosity. 
Generally, the conclusion is that longer alkyl chains lead to better tribological performance.
Related to the impact of alkyl chain length on the structure of ILs, \citeauthor{perkin2011self}~\cite{perkin2011self} have experimentally obtained the formation of tail--to--tail bilayers of cations
if their alkyl chain length is large, in case of confinement between solid surfaces. 
Their observations are in accordance with other experimental investigations of IL lubricants~\cite{wang2004friction,jimenez20061,mu2005effect}. In this work, we have obtained similar configurations via numerical simulations of ILs confined between two solid plates, where tail--to--tail formation in the middle of the interplate gap is visible.

In this theoretical study, we apply a coarse grain Molecular Dynamics (MD) simulation setup consisting of two solid plates and an IL placed between them. 
Our simulation setup also includes lateral reservoirs into which the IL can dynamically expand~\cite{tribint2017}. 
The focus of our study is on the systematic investigation of the flow properties and lubrication mechanisms of ionic liquids modelled with a generic coarse grain model which considers a variable shape of the cation. 
We investigate the impact of cationic tail size on the structural and tribological properties of ILs via molecular dynamics simulations. 
Such an idea is meaningful since previous theoretical studies have pointed out that confinement modifies the behaviour of ILs and despite the good wetting nature, the slip is present at the plates~\cite{VoeltzelC5CP03134F}. 
Coulombic interactions in ILs induce long-range ordering~\cite{mendonca2013ILmetal,VoeltzelC5CP03134F,CanovaC4CP00005F}, which in turn can influence their lubrication response. 
Currently, there is a substantial modelling effort towards the investigation of ILs as lubricants~\cite{fajardo2015friction,fajardo2015electrotunable,capozza2015squeezout}. 
Coarse grain approaches, being less computationally expensive, have an advantage for reaching the length-- and time-- scales that can be of relevance to the systems of industrial interest. 
Previously, coarse grain MD simulations~\cite{GaoJPCB2004,Robbins2000,PhysRevB.58.R5893,ar700160p,Heyes1.3698601,GattinoniPhysRevE.88.052406,Martinie2016} were used to study thin lubricant films subjected to
the shearing between solid plates.

We outline the content of this paper: \textit{Model} section describes the interactions taken into account and the MD simulation setup. 
The focus of \textit{Bulk Ionic Liquid} section is first on obtaining the relaxed structure and then on calculating the viscosity coefficient of bulk ionic liquids. 
In the following \textit{Confined Ionic Liquid} section we present and discuss the static and dynamic behaviour of confined ionic liquids.  
This section also presents the results of confined IL's friction behaviour. We presents the overview of contributions in \textit{Discussion} section followed by the \textit{Conclusion}.

\vspace{0.6cm}
%%%%%%%%%%%%%%%%%%%%%%
%FIG. 1
\begin{figure}[h!]
\includegraphics[width = 8 cm]{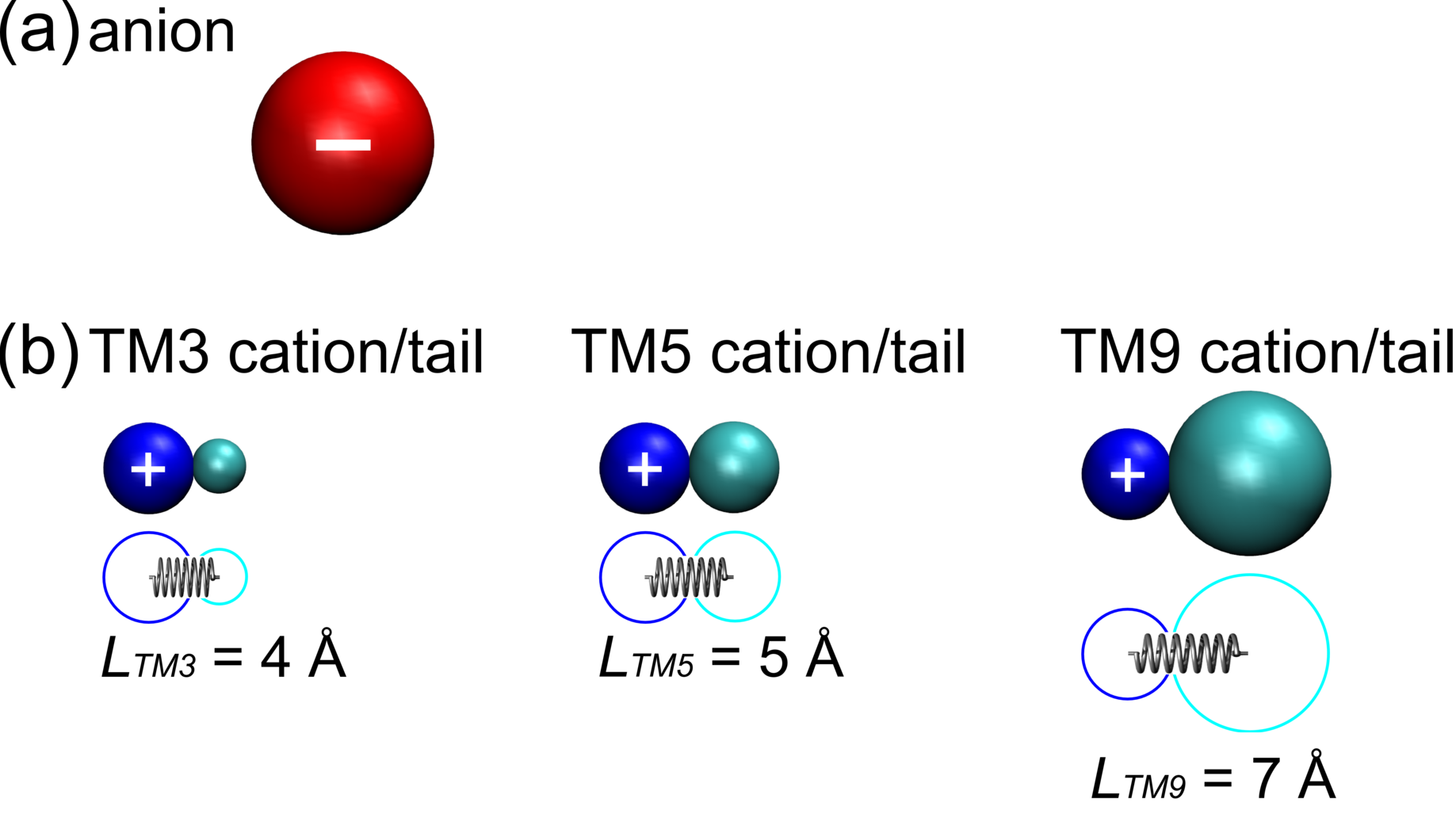}
\caption{Schematic representation of (a) anion and (b) cation molecules in TM model. The anion is represented by a spherical particle with a diameter $\sigma_{\rm A}=10$~\AA. The cation molecule consists of a charged head with a diameter $\sigma_{\rm C}=5$~\AA and a neutral tail. In order to be more concise, we refer just to cationic head as the cation. The cation and its tail are connected by a spring with length $L=(\sigma_{\rm C}+\sigma_{\rm T})/2$. The size of the tail has been varied and (a) TM3, (b) TM5 and (c) TM9 ionic liquids have a tail diameter of $3$, $5$ and $9$~{\AA}, respectively. The molecular asymmetry is a feature of real ionic liquids and chosen parameters resemble $\left[BMIM\right]^{+}\left[PF_6\right]^{-}$ IL properties, cf. Ref.~\cite{fajardo2015friction,fajardo2015electrotunable}.
}
\label{fig:ILmodel}
\end{figure}
%%%%%%%%%%%%%%%%%%%%%%
\section{Model}\label{Model}
In this study, we have applied a generic coarse grained IL model, introduced in Ref.~\cite{capozza2015squeezout}. In this model, the anion is represented as a negatively charged large--sized spherical particle, while the cation is a dimer consisting of a positively charged small--sized spherical particle (i.e. cationic head),
and a neutral spherical particle (tail) attached to the corresponding cationic head via an elastic spring, see Figure~\ref{fig:ILmodel}. Since the cationic tail is the principal feature of the model used in this paper, we will refer to it as tail model (TM). The asymmetry of the cation leads to amorphous (glassy) states for realistic values of interaction parameters (e.g., for hydrocarbons), in contrast to the simplest coarse--grained model of IL known as SM model (salt--like model), where both cations and anions are spherical.  The SM model has already been exploited in previous studies~\cite{fajardo2015friction,capozza2015squeezout,tribint2017, DasicEPJE2018}. Despite an obvious advantage of simplicity, in order to avoid crystallization, the SM model relies on a very weak non-bonded Lennard-Jones interaction which makes any comparison with real IL only qualitative. In addition, the SM model cannot account for molecular asymmetry featured in real ILs. Nevertheless, the SM model has been proven to be quite useful for the development of the simulation methodology, as it reduces computational complexity and enables faster equilibration (e.g., for obtaining static force distance characteristics as in Ref.~\cite{tribint2017}). More complex extensions of TM coarse grain models can involve several tails of different size, like in Ref.~\cite{fajardo2015friction}. For simplicity reasons, we restrain our considerations in this study to a single neutral tail of a variable size. Although a whole cationic dimer is an entity which actually represents a cation, in order to be more concise we refer just to cationic head as the cation. 

%One might raise a question what are the reasons for the attaching of a neutral tail to a cation? First of all, real ILs usually include cations that consist of the cationic head (positively charged) and alkyl chain (neutral part of cation). Alkyl chains can have different lengths (different number of $C$ atoms). Furthermore, the tail enhances the general tendency of ILs to form a glass rather than a crystal at low temperatures~\cite{capozza2015squeezout}.  As the previous studies have shown, the shape of IL molecules may affect their layering structure~\cite{fajardo2015friction}.  According to that, the central question which we address in this study is how does the tail size affect the structure, static and dynamic behaviour, as well as, lubrication properties of a generic IL represented via tailed--model. 

%%%%%%%%%%%%%%%%%%%%%%
%FIG. 2
\begin{figure*}[ht!]
\includegraphics[width = 15 cm]{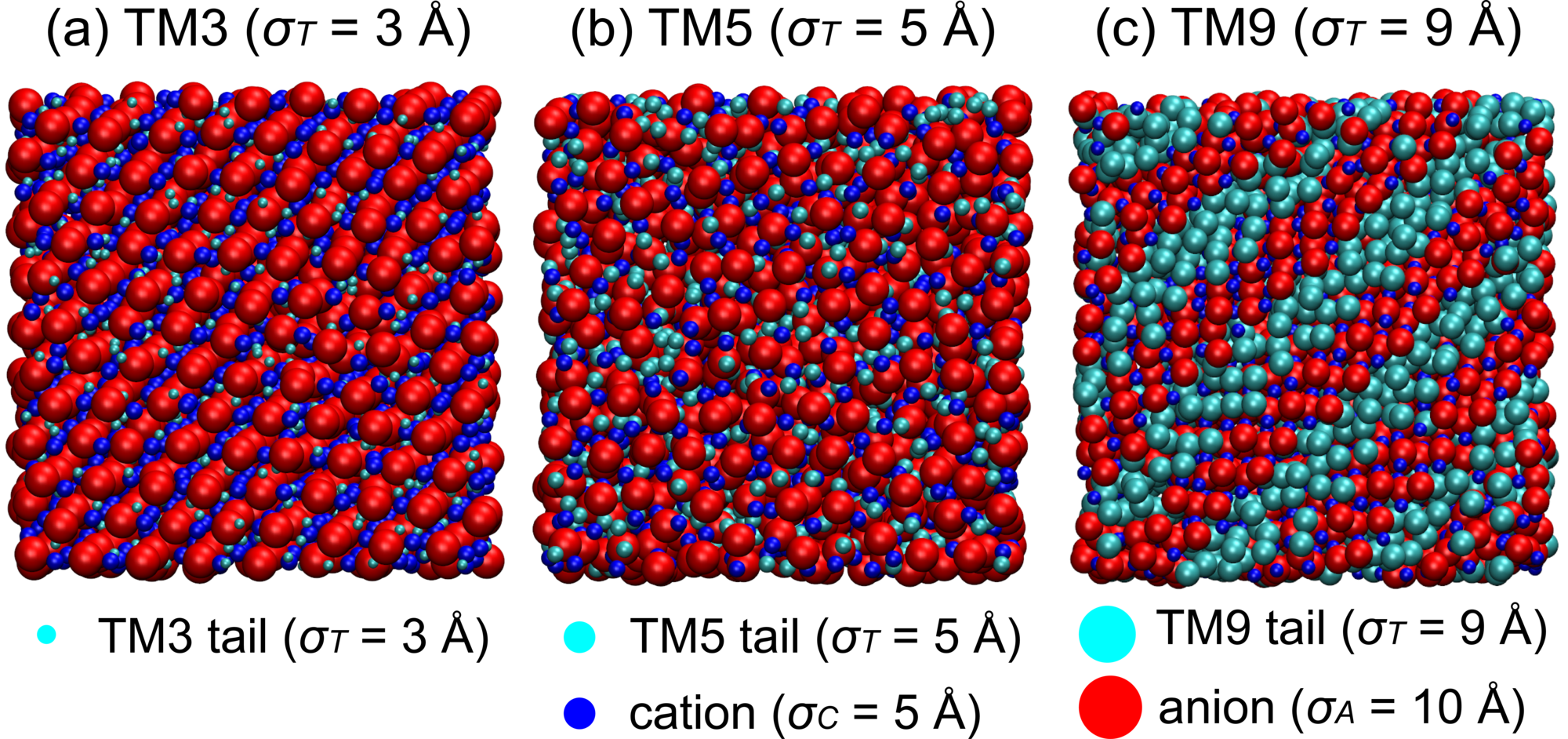}
\caption{Configuration snapshots of bulk (a) TM3, (b) TM5 and (c) TM9 ionic liquid, with tail diameter of $3$, $5$ and $9$~{\AA}, respectively. 
We may notice that each configuration snapshot represents a different state, i.e. TM3 bulk IL crystallizes into
a tilted simple cubic crystal structure, oriented along the face diagonal; TM5 bulk IL is in liquid state; TM9 bulk IL crystallizes into crystal planes with alternating
ionic--tail layers, oriented along the face diagonal as well.
}
\label{fig:bulkconfTM}
\end{figure*}
%%%%%%%%%%%%%%%%%%%%%%

\subsection{Interaction Model}
In cation--tail dimers an elastic spring connects cations and neutral tails enabling the tail's freedom of moving independently from its cation, since their connection is not rigid, cf. Figure~\ref{fig:ILmodel}.    
Interatomic interactions taken into consideration in our MD simulations are: $(i)$ non--bonded  Lennard--Jones (LJ) and Coulombic electrostatic interactions, 
and $(ii)$ bonded interaction (elastic spring potential in cation--tail pairs):
\begin{equation}
V_{\alpha\beta}\left(r_{ij}\right) =  4 \epsilon_{\alpha\beta}
\left[\left(\frac{\sigma_{\alpha\beta}}{r_{ij}}\right)^{12}-
\left(\frac{\sigma_{\alpha\beta}}{r_{ij}}\right)^6\right] + \frac{1}{4 \pi \epsilon_0 \epsilon_r} \frac{q_iq_j}{r_\mathrm{\textit{ij}}},
\end{equation}
where $\textit{i},\textit{j}=1,\dots,N$ are particle indices, and $N$ is the total number of particles. Particles can be of different types $\alpha,\beta = {{\rm A}, {\rm C}, {\rm T}, {\rm P}}$ which refer to anions, cations, tails, and solid plate atoms, respectively. Interaction of tails (i.e., at least one of indices $\alpha,\beta = {\rm T}$) with all other atom types, including tails themselves, is implemented using a purely repulsive potential.
%\begin{equation}
%V_{\alpha\beta}\left(r_\mathrm{\textit{ij}}\right)=\epsilon_{\alpha\beta} + 
%4 \epsilon_{\alpha\beta}
%\left[\left(\frac{\sigma_{\alpha\beta}}{r_{ij}}\right)^{12} -
%\left(\frac{\sigma_{\alpha\beta}}{r_{ij}}\right)^6\right],~ %r_{ij}\leq 2^{1/6}\sigma_{\alpha\beta}
%\end{equation}
%and $V^{\rm LJ}\left(r_\mathrm{\textit{ij}}\right)=0,~r_{ij}>2^{1/6} \sigma_{\alpha\beta}$. 
The ionic liquid is electro--neutral, i.e., the number of cations and anions is the same.  
%The total number of ionic liquid molecules (cation--tail dimers and anions) is $N_{\rm IL} = 3000$. Therefore, the total number of ions is $N_{\rm C} = N_{\rm A} = 1000$ and the number of tails is $N_{\rm T} = N_{\rm C} = 1000$.
All MD simulations in this study were performed using the LAMMPS software~\cite{plimpton1995fast}. More details are provided in the Suplementary Information (SI).

\subsection{Model Parameters}
In this study we have fixed the diameter of the cationic heads and anions to $\sigma_{\rm C} = 5$~{\AA} and $\sigma_{\rm A} = 10$~{\AA}, respectively. 
Such choice respects the asymmetry that exists in ILs and it is consistent with other models, as well as, for example $\left[BMIM\right]^{+}\left[PF_6\right]^{-}$ ionic liquid, cf. Ref.~\cite{fajardo2015friction,fajardo2015electrotunable,capozza2015squeezout,tribint2017}.   
The solid plate atoms have a diameter of $\sigma_{\rm P} = 3$~{\AA}.
We have taken into consideration three different tailed-models of IL depending on the tail size, which is defined by its Lennard--Jones $\sigma_{\rm T}$ parameter: 
small--tail cationic dimer (i.e., TM3 model with $\sigma_{\rm T} = 3$~{\AA}), symmetric cationic dimer (i.e., TM5 model with $\sigma_{\rm T} = \sigma_{\rm C} = 5$~{\AA}) 
and large--tail cationic dimer (i.e., TM9 model with $\sigma_{\rm T} = 9$~{\AA}), see Figure~\ref{fig:ILmodel}.
Drawing a comparison with the experiment in Refs.~\cite{dold2013influence,minami2009ionic}, the TM~IL mimics a folded alkyl chain and the radius of the sphere is related to the gyration radius of the chains. Depending on the length of the alkyl chain, the sphere has a smaller or lager radius. 
Thus, the size of a sphere which represents a neutral tail in TM~ILs does not compare directly with the alkyl chain length. However, we can make a qualitative analogy. 
While the representation of the alkyl chain as a neutral LJ sphere does not include all the microscopic level features,
we will show that the three selected radii, i.e., $\sigma_{\rm T} = \left\{3, 5, 9\right\}$~{\AA}, result in clear differences of the bulk properties of ILs and their lubrication response.

Each cation--tail pair is connected via an elastic spring defined by the next two parameters:
elastic constant $K = 80$~$\text{kcal}/\text{mol}\text{{\AA}}^2$ and equilibrium length of the spring $L = \left(\sigma_{\rm C} + \sigma_{\rm T}\right)/2$. 
To account for the dielectric screening, the dielectric constant is set to $\epsilon_r = 2$ as in Refs.~\cite{fajardo2015electrotunable, capozza2015squeezout, tribint2017}.  
The strength of the LJ interactions between different charged parts of ions ($\alpha, \beta = {{\rm A}, {\rm C}}$) is $\epsilon_{\rm \alpha\beta} = 1.1$~kcal/mol. The LJ parameters are chosen to compare well with one of the most widely studied ionic liquids $\left[BMIM\right]^{+}\left[PF_6\right]^{-}$, cf. Ref.~\cite{fajardo2015friction,fajardo2015electrotunable}. 
The charge of ions is set to elementary: $q_{\rm C} = + \it{e}$ and $q_{\rm A} = - \it{e}$, where $\it{e} = \text{1.6} \cdot \text{10}^{\text{-19}}$~C. 
The tails interact with all other particle types repulsively. 
The strength of the ion-substrate interaction was tuned to ensure complete wetting, $\epsilon_{\rm \alpha P} = 5.3$~kcal/mol, $\alpha = {{\rm A}, {\rm C}, {\rm T}}$.~\footnote{Only when the strength of ion-substrate LJ interaction equals the strength of inter-ionic LJ interaction, partial wetting is observed, i.e., $\epsilon_{\rm \alpha P} = 1.1$~kcal/mol, as reported in the SI.} All the values of $\{\epsilon_{\rm \alpha\beta}, \sigma_{\rm \alpha\beta}\}$ parameters used in our simulations are listed in SI. 
Cross-interaction parameters are calculated by Lorentz-Berthold mixing rules.

\vspace{0.6cm}
\section{Bulk Ionic Liquid}\label{Bulk}

\subsection{Bulk structure}
An initial configuration for bulk ionic liquid was obtained by a random placement of ions ($N_{\rm C} = N_{\rm A} = 1000$) into a cubic simulation box with periodic boundary conditions in all three directions. 
The simulation box volume was chosen to ensure that the resulting pressure after the relaxation of the IL structure is comparable to the one experienced by thin confined IL film studied in the following section of this paper.  
In the case of the present system the pressure is $p \approx 10$~MPa, which corresponds to a normal force of 1~nN acting on a surface of $10^4$~\AA$^2$. Relaxation of the internal energy and pressure for the three TM~ILs is presented in detail in the SI.

Figure~\ref{fig:bulkconfTM} presents the $xy$ cross--section snapshots of bulk IL configurations at the end of relaxation simulations, again for (a) TM3, (b) TM5 and (c) TM9 model. These results have clearly revealed a strong dependence of IL's structure on the tail size. It can be observed that $(i)$ small tails in the TM3 model lead to cubic crystalline arrangement of ions, $(ii)$ symmetric cationic dimers in the TM5 model enable a liquid-like state of IL, and $(iii)$ large tails in the TM9 model dictate ordering in the way that ions form layers with tails in-between. 

These results are in agreement with experimental observations of the relation between the length of alkyl chain and the structure of bulk IL~\cite{hayes2015structure}. When the cation alkyl chain is short, the Coulombic forces are dominant, enabling order. We observe this kind of result with TM3 model.
Alkyl chain must be long enough in order to suppress the Coulombic interactions, e.g. number of C atoms $n_{\rm C} \approx 12$, which corresponds to tail length of $(n_{\rm C} - 1) \cdot 1.53~\text{\AA} = 16.83~\text{\AA}$, taking into account
that a C-C bond has a length of $1.53~\text{\AA}$. 
Suppressing of Coulombic interactions results in the absence of order, as we obtain with TM5 model, cf. Figure~\ref{fig:bulkconfTM}(b). However, the tail should not be too large since large tails tend to arrange into a separate layer. This leads to a reappearance of layered structural ordering, like in the case of TM9 model,  cf. Figure~\ref{fig:bulkconfTM}(c). This layering can take place even when the cohesive interaction between the tails is absent, since in our TM~IL the pair-interaction of tails with all other particles is repulsive.

%%%%%%%%%%%%%%%%%%%%%%
%FIG. 3
\begin{figure}[th!]
\includegraphics[width = 7 cm]{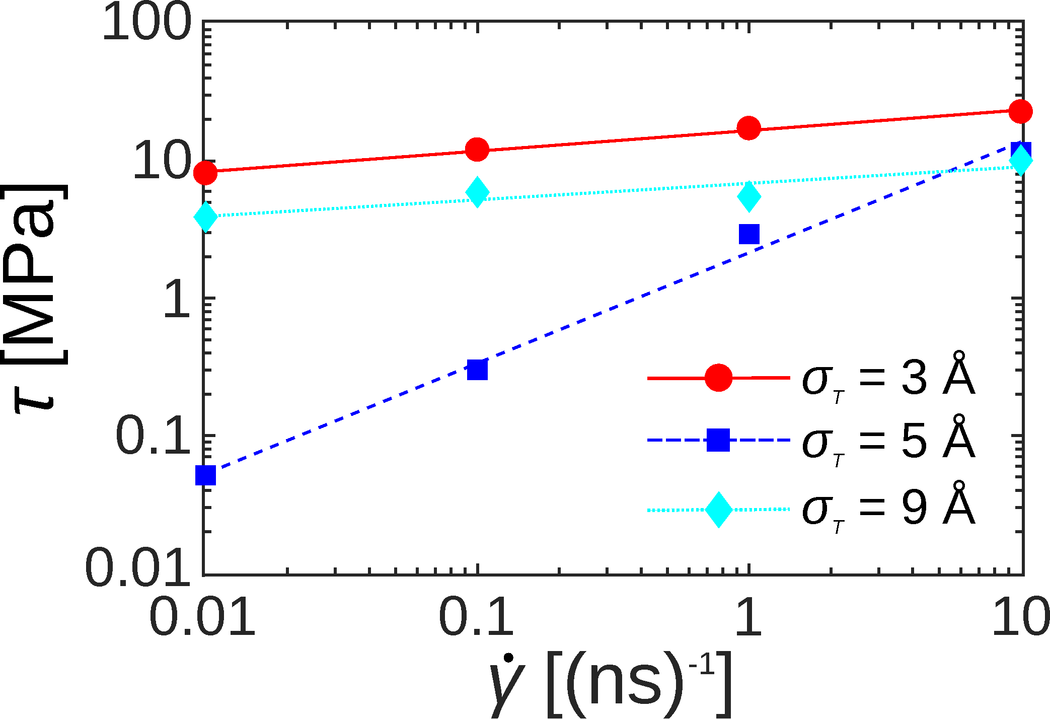}
\caption{Average shear stress $\tau$ as a function of shear rate $\dot{\gamma}$ of TM3, TM5 and TM9 bulk IL.
We have conducted shear simulations for the shear rates in a range of four orders of magnitude ($\dot{\gamma}=0.01 - 10$~$\text{ns}^{\text{-1}}$). 
The lines are obtained by fitting the points with Eq.~\ref{eqn:tau_gamma}. 
} 
\label{fig:tau_shear}
\end{figure}
%%%%%%%%%%%%%%%%%%%%%%

\subsection{Bulk IL viscosity characteristics}
We have calculated the viscosity using non--equilibrium (NEMD) simulations of the three TM~IL systems in a box with periodic boundary conditions in all three directions under different shear rates. Since in bulk simulations the whole simulation box is sheared, the shape of the box changes. Therefore, we use the so-called \textit{SLLOD} thermostat~\cite{evans1984nonlinear,daivis2006simple} (more details are provided in the SI).  For each value of $\dot{\gamma}$ in the range $0.01 - 10$~$\text{ns}^{\text{-1}}$, we calculate the average shear stress from three stress tensor components: $\tau = \left(\tau_{xy} + \tau_{xz} + \tau_{yz}\right)/3$. The average shear stress $\tau$ and shear rate $\dot{\gamma}$ are related by 
\begin{equation}
\tau = \eta \cdot \dot{\gamma}^{\alpha},
\label{eqn:tau_gamma}
\end{equation} where $\eta$ is the generalised viscosity coefficient and $\alpha$ is an exponent.  In addition to the NEMD method of simulation box shearing, we have also calculated the zero shear rate  viscosity $\eta^{\rm GK}$ using the Green-Kubo (GK) relation for the three model ILs, as the integral of the stress tensor auto-correlation functions, see Ref.\cite{green,kubo}.

%The condition which ensures the applicability of equation~\ref{eqn:tau_gamma} to liquids is: $\alpha \approx 1$. This is a numerical condition corresponding to the liquid state of the "liquid" under investigation. Therefore, the equation~\ref{eqn:tau_gamma} does not apply to those cases, hence we do not calculate their viscosity coefficients $\eta$. The results of shearing simulations are consistent with the results of relaxation simulations, since they indicate that TM3 and TM9 bulk IL are crystalized and that TM5 bulk IL is in a liquid state. We might make an observation that the shearing simulations did not show themselves as quantitatively precise, but as a useful probing tool from the qualitative point of view. In principle, Figure~\ref{fig:tau_shear} is the confirmation that TM5 model of bulk IL exhibits viscosity characteristics typical for a liquid, while TM3 and TM9 models do not, which is in agreement with the results of the relaxation simulations. 

In Figure~\ref{fig:tau_shear} we present the dependence of the average shear stress $\tau$ on the shear rate $\dot{\gamma}$ for TM3, TM5 and TM9 bulk IL. 
We notice that the average shear stress stays within the same order of magnitude in TM3 and TM9 systems, although the shear rate changes four orders of magnitude. As a result, the corresponding values of the exponent $\alpha$ are low, i.e. $\alpha_{\rm TM3} = 0.15 \pm 0.02$ and $\alpha_{\rm TM9} = 0.12 \pm 0.04$. The bulk IL in the case of TM3 and TM9 models is ordered. The presence of order results also in high values of their Green-Kubo viscosities, i.e. $\eta^{\rm GK}_{\rm TM3} = 4.72$~$\text{mPa} \cdot \text{s}$ and $\eta^{\rm GK}_{\rm TM9} = 1.67$~$\text{mPa} \cdot \text{s}$.
In contrast to that, we observe more than two orders of magnitude change of the average stress tensor component in the case of symmetric cations and liquid-like bulk structure (TM5 model). We have obtained $\alpha_{\rm TM5} = 0.8 \pm 0 .1$, which is relatively close to a Newtonian viscous fluid, i.e., $\alpha = 1$.
The viscosities determined via shearing simulations and via GK relation in case of TM5 model are different, however they are of the same order of magnitude:
$\eta_{\rm TM5} = 0.1435$~$\text{mPa} \cdot \text{s}$ and $\eta^{\rm GK}_{\rm TM5} = 0.6144$~$\text{mPa} \cdot \text{s}$, respectively. 

%\begin{table}
%\begin{center}
%\begin{tabular}{ |c|c|c|c| }
% \hline
% IL & $\eta$ [$\text{mPa} \cdot \text{s}^{\alpha}$] &  $\alpha$ & %$\eta^{\rm GK}$ [$\text{mPa} \cdot \text{s}$] \\
% \hline
% TM3 & 4.72 & 0.15 $\pm$ 0.02 & $30$ \\
% \hline
% TM5 & 0.1435 & 0.8 $\pm$ 0.1 & 0.6144  \\
% \hline
% TM9 & 1.67 & 0.12 $\pm$ 0.04 & $30$  \\
% \hline
%\end{tabular}
%\end{center}
%\caption{Overview of the results of relaxation simulations: $\sigma_{\rm TT}$ is the tail size, $L$ is the side length of cubic simulation box, $t_{\rm rel}$ is the estimated relaxation time, 
%$\left\langle p \right\rangle$ and $\left\langle E_{\rm int} \right\rangle$ are the mean values of pressure and internal energy respectively, averaged over the time span $t_{\rm rel} \leq t \leq t_{\rm tot}$, 
%where $t_{\rm tot}$ is the total simulation time. 
%}
%\label{tab:bulkTM}
%\end{table}

%%%%%%%%%%%%%%%%%%%%%%
%FIG. 5
\begin{figure}[ht!]
\includegraphics[width = 7 cm]{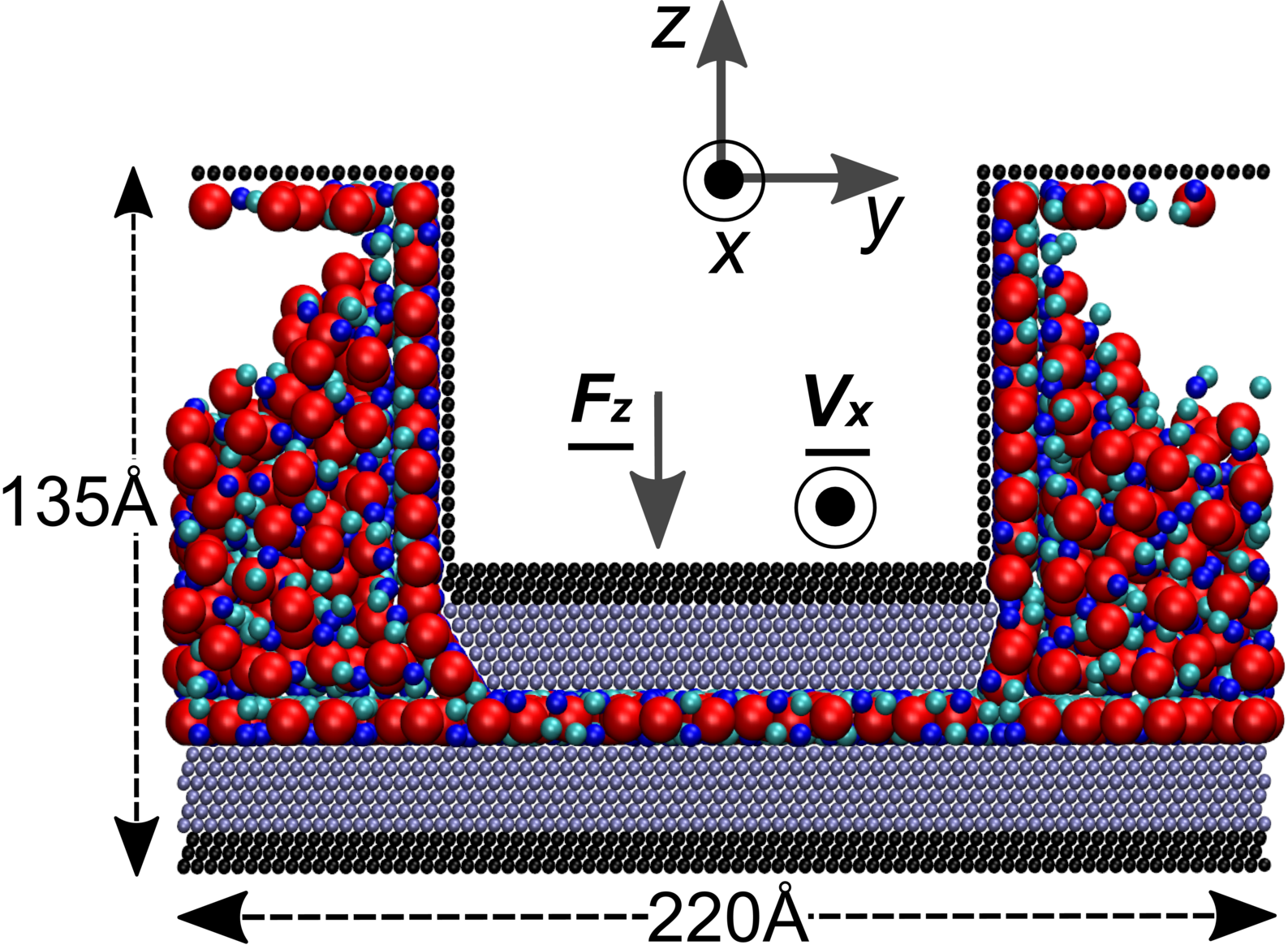}
\caption{Schematic of the simulation setup shown as $yz$ cross-section. 
Dimensions of the system along the $y$ and $z$ axes, together with the directions of the imposed normal load $F_z$ and lateral velocity $V_x$ are noted. The total system length in the $x$ direction is $125$~{\AA}.  
There are two solid plates at the top and bottom of the system (more details on the simulation configuration is given in SI). In the coloured version of the paper the different regions have different colours. 
The ionic liquid is composed of an equal number of cation--tail pairs and anions (in the coloured version of the paper particles can be visually distinguished: cations - blue spheres, tails - cyan spheres, and anions - red spheres).
}
\label{fig:schemeTM}
\end{figure}
%%%%%%%%%%%%%%%%%%%%%%

%%%%%%%%%%%%%%%%%%%%%%
%FIG. 6
\begin{figure}[ht!]
\includegraphics[width = 8 cm]{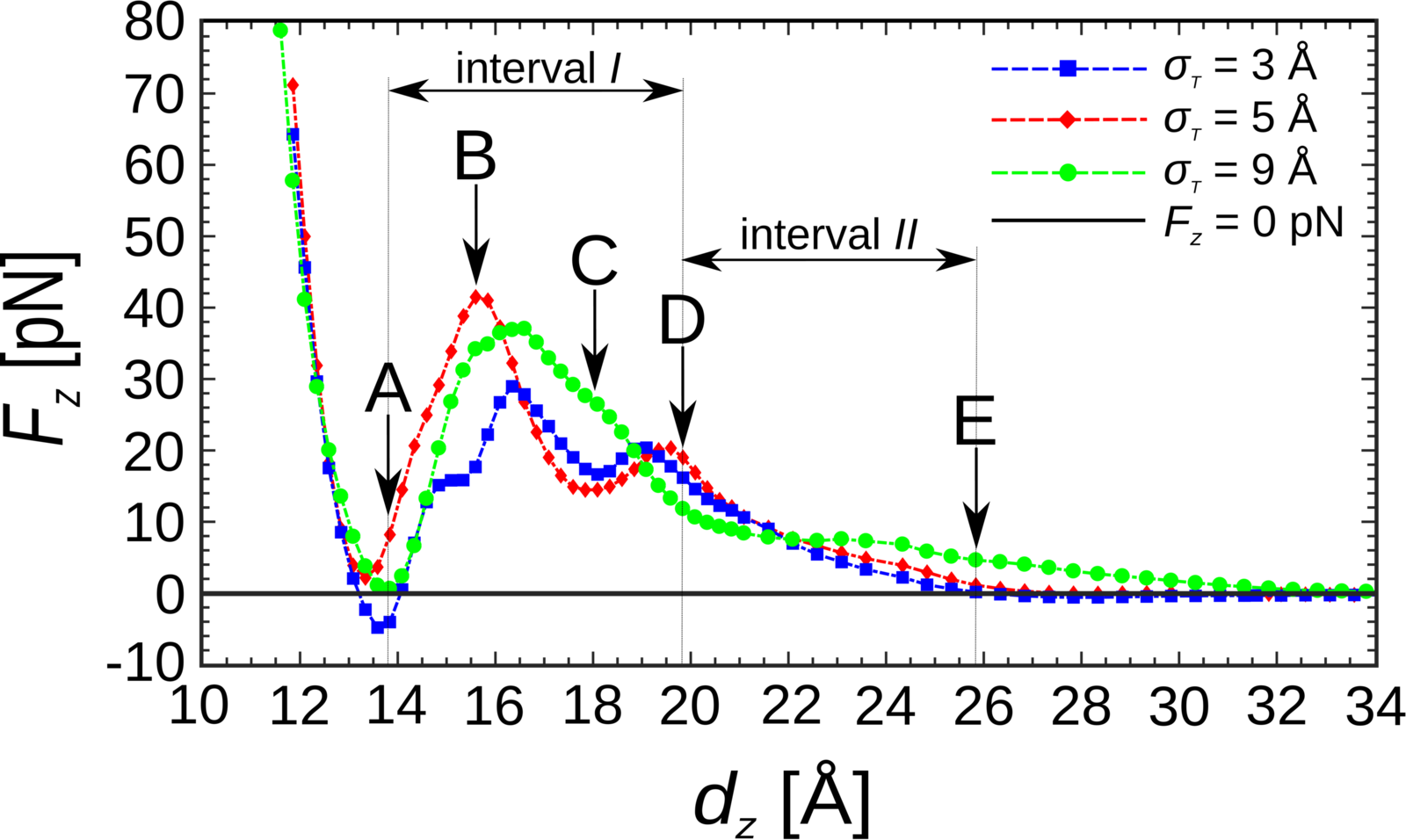}
\caption{Dependence of normal force $F_z$ on plate-to-plate distance $d_z$.
Five characteristic points denoted with \{A, B, C, D, E\} with corresponding interplate distances
$d_z = {13.8, 15.5, 18.0, 19.8, 25.8}$~{\AA}, respectively, are marked in the figure. 
They are chosen in the way that:
point A is located in the proximity of a local minimum for all three cases;
point B corresponds to a local maximum for TM5 model; 
point C is located in the proximity of a local minimum for TM3 and TM5 model;
point D is located in the proximity of a local maximum for TM3 and TM5 model; 
point E is chosen according to the condition $\overline{DE} = \overline{AD}$. 
For reference, the black horizontal line denotes $F_z = 0$. The lines connecting points (averages of normal force) serve as visual guide. 
}
\label{fig:fd_static}
\end{figure}
%%%%%%%%%%%%%%%%%%%%%%

%%%%%%%%%%%%%%%%%%%%%%
%FIG. 8
\begin{figure}[ht!]
\includegraphics[width = 7 cm]{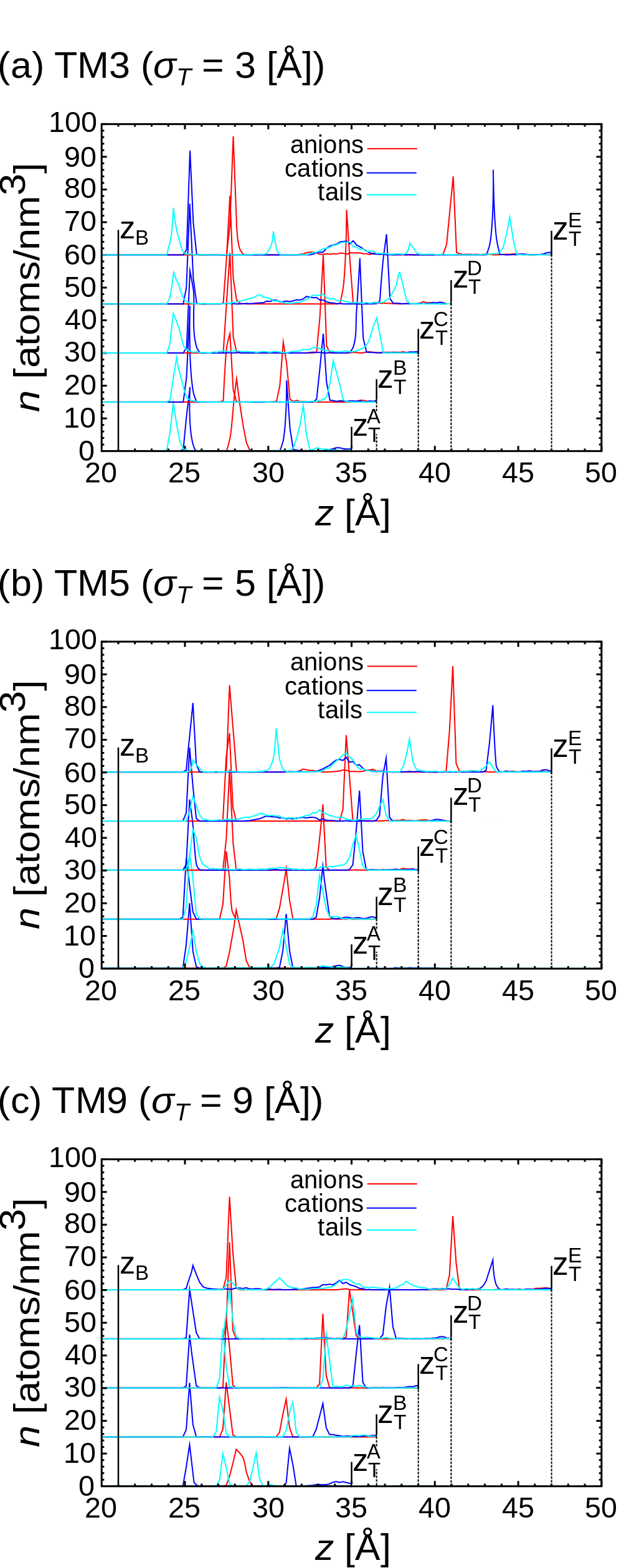}
\caption{Ionic density distribution of ions inside the interplate gap of (a) TM3, (b) TM5 and (c) TM9 models in characteristic points $\left\{A, B, C, D, E\right\}$ selected in the static force--distance characteristic presented in the Figure~\ref{fig:fd_static}. The positions of the atomic centres of the innermost atomic layers of the (moving) top and (fixed) bottom plate are labeled as $z_{\rm T}^{A-E}$ and $z_{\rm B}$, respectively. Five characteristic points, denoted with \{A, B, C, D, E\}, have corresponding interplate distances
$d_z=z_{\rm T}-z_{\rm B}= {13.8, 15.5, 18.0, 19.8, 25.8}$~{\AA}, respectively.
}
\label{fig:ionic_distribution}
\end{figure}
%%%%%%%%%%%%%%%%%%%%%%

%%%%%%%%%%%%%%%%%%%%%%
%FIG. 7b
%\begin{figure*}[t!]
%\includegraphics[width = 18 cm]{fig7b_resized.png}
%\caption{Configuration snapshots ($yz$ cross section) of TM3, TM5 and TM9 models in five characteristic points $\left\{A, B, C, D, E\right\}$. 
%This figure gives changes taking place in the confined ionic layers as the interplate distance changes in case of static force--distance simulations. Five characteristic points, denoted with \{A, B, C, D, E\}, have corresponding interplate distances
%$d_z = {13.8, 15.5, 18.0, 19.8, 25.8}$~{\AA}, respectively.
%}
%\label{fig:confyz_b}
%\end{figure*}
%%%%%%%%%%%%%%%%%%%%%% 

\section{Confined Ionic Liquid}
\label{Confined}
For the study of IL under confinement, we use the MD simulation setup of ILs under confinement shown in Figure~\ref{fig:schemeTM}. The ionic liquid is placed between two solid plates: a {\it bottom plate} which is continuous in two dimensions (in $xy$-plane) and a {\it top plate} which is infinite in one dimension (along the $x$-axis) and features lateral reservoirs in the other, i.e., along the $y$-axis. This design allows a long-range ordering of the ILs on the surface while at the same time creating quasi micro-canonical conditions inside the interplate gap. We use this setup throughout the paper in order to investigate both the static and dynamic behaviour of the confined IL, as well as, its lubrication performance. We keep the simulation setup geometry fixed, and we change the IL.
Additional implementation details can be found in the SI. 

\subsection{Equilibrium Behaviour of Confined Ionic Liquid}
Confinement induces layering in IL thin films~\cite{PerkinPCCP2012,tribint2017}. In order to understand how does an interplay between layering and molecular geometry of IL alter the load bearing capability of the thin films, we calculate the quasi-static force-distance characteristic. We follow the evolution of the normal load $F_z$ acting on the top plate as a function of interplate distance $d_z$. To ensure static conditions, the interplate distance is changed through a series of alternating steps, called \textit{move} and \textit{stay}, related to the movement of the top plate and subsequent relaxation of the IL structure, respectively. We describe in detail the simulation procedure in the SI. The results for the force-distance characteristic of the three TM ILs are presented in Figure~\ref{fig:fd_static}, where three different markers correspond to the three IL models. The normal force $F_z$ strongly and non-monotonously depends on the distance $d_z$. These changes of the normal force $F_z$ are correlated with the squeezing in and out of cation/anion layer pairs into the gap, as already observed experimentally~\cite{hayes2011double} and theoretically~\cite{tribint2017}. The normal force becomes negative ($F_z<0$) only in the case of small tails (TM3 model). The negative values are a result of the IL trying to reduce the plate-to-plate distance due to the adhesion forces inside of IL. The increasing tail size seems to reduce the effect of adhesion: for large tails (TM9) the normal force at the minimum is close to zero, while for symmetric cation molecule (TM5) it becomes positive ($F_z=2$pN).

%%%%%%%%%%%%%%%%%%%%%%
%FIG. 8
\begin{figure}[ht!]
\includegraphics[width = 7 cm]{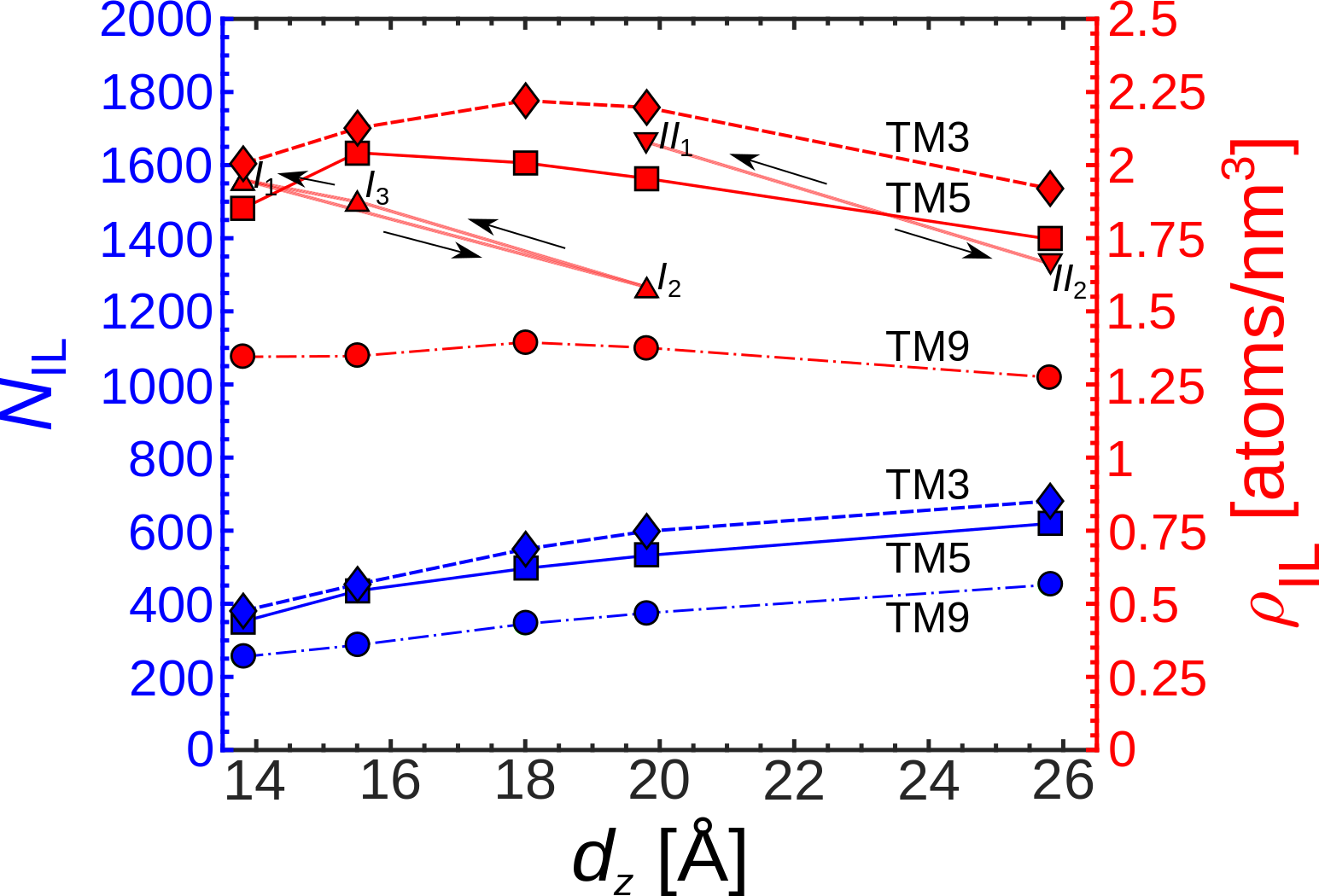}
\caption{Evolution of the number of confined ionic liquid (IL) molecules (bottom curves) and density (top curves) inside the gap with gap width $d_z$ for TM3, TM5 and TM9 models in characteristic points $\left\{A, B, C, D, E\right\}$ selected from the static force--distance characteristic 
(Figure~\ref{fig:fd_static}). The corresponding axes for the number of IL molecules and the density are given on the left and right side, respectively. The densities at characteristic points for the dynamic cases (intervals $I,II$) are also given, i.e., $I_{1,2,3}$ and $II_{1,2}$. Five characteristic points denoted with \{A, B, C, D, E\} in static and $I_{1,2,3}$ and $II_{1,2}$ dynamic case have the same corresponding interplate distances $d_z = {13.8, 15.5, 18.0, 19.8, 25.8}$~{\AA}, respectively.}
\label{fig:ionic_density}
\end{figure}
%%%%%%%%%%%%%%%%%%%%%%

%%%%%%%%%%%%%%%%%%%%%%
%FIG. 9
\begin{figure}[ht!]
\includegraphics[width = 7 cm]{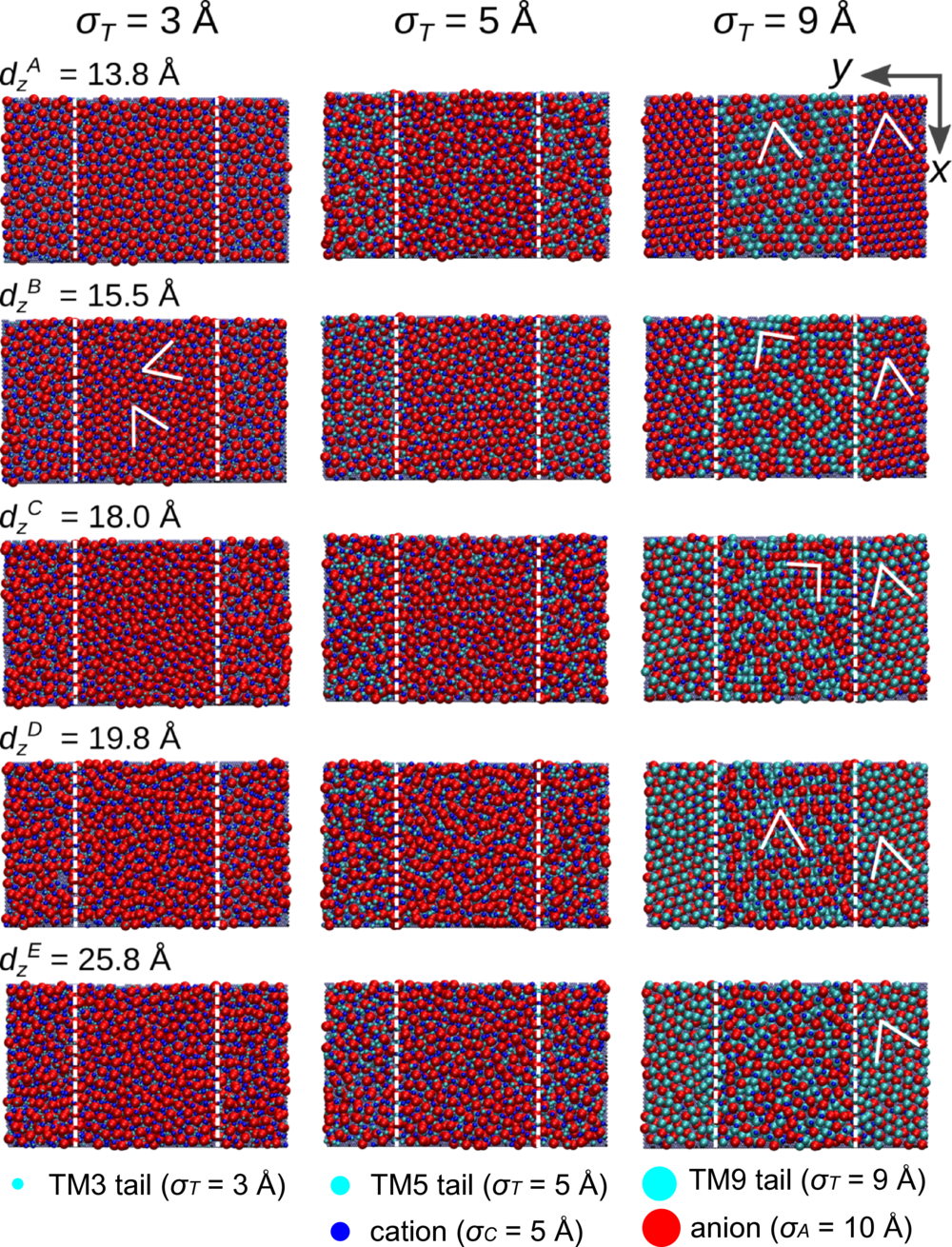}
\caption{Configuration snapshots ($xy$ cross section) of TM3, TM5 and TM9 models in five characteristic points $\left\{A, B, C, D, E\right\}$. Five characteristic points, denoted with \{A, B, C, D, E\}, have corresponding interplate distances
$d_z = {13.8, 15.5, 18.0, 19.8, 25.8}$~{\AA}, respectively (see also Figure~\ref{fig:fd_static}).
}
\label{fig:confxy}
\end{figure}
%%%%%%%%%%%%%%%%%%%%%%

For all three curves corresponding to the three TM~ILs we can identify three characteristic ranges of the plate-to-plate distance $d_z$: initial segment ($11$~{\AA}$ \leq d_z \leq 13.8$~{\AA}) characterized by a monotonous and steep decrease of the normal force $F_z$; interval \textit{I} ($13.8$~{\AA}$ \leq d_z \leq 19.8$~{\AA}) characterized by the presence of local minima and maxima peaks of the normal force $F_z$ , and interval \textit{II} and beyond ($d_z \geq 19.8$~{\AA}) characterized by a continuous and gentle decrease of the normal force $F_z$, where in all three cases the normal force practically becomes zero when $d_z > 32$~{\AA}. 

We will briefly describe the segments of $F_z\left(d_z\right)$ curves, pointing out similarities and differences between the different IL models.  
In the initial segment, i.e., for small gaps $d_z < 13$~{\AA}, the normal force $F_z$ is practically the same for all three systems, meaning that it does not depend on the tail size. 
The steep rise of the normal force with compression in the range $d_z < 13$~{\AA} is a sign of very high resistance of the single anionic layer left in the gap to squeeze out.
On the other hand, at large gap values (i.e., $d_z > 32$~{\AA}), the normal load $F_z$ in all three TM~ILs is similar and small. We can conclude that at large gaps there is a low resistance of IL to the gap changes. Significant differences in the force-distance curves depending on the tail size exist only in the interval \textit{I}, i.e., $13.8$~{\AA}$ \leq d_z \leq 19.8$~{\AA}.
In the case of the TM3 model, the $F_z\left(d_z\right)$ characteristic has two local minima and maxima and one saddle point, in the TM5 model there are two local minima and maxima, and
for the TM9 model, there is one local minimum and maximum. 

% We might label the layers formed alongside the solid plates as fixed layers, since they always form first. 

%Inside the interplate gap ionic ordering is dictated according to the layers formed next to the solid plates: $(i)$ bottom plate - cation--tail layer anionic layer, looking from the bottom,
%$(ii)$ top plate - cation--tail layer - anionic layer, looking from the top, where bottom and top correspond to the position along the $z$ axis. 
%In Figures~\ref{fig:confyz_b} and~\ref{fig:confxy} we present $5 \times 3$ panels of configuration snapshots for $5$ chosen characteristic points of $3$ TM models. Atoms are depicted keeping the ratios of their sizes. 

\subsubsection{IL layer structure inside the gap}
In Figure~\ref{fig:ionic_distribution} we are showing the ionic density distribution along the $z$ axis for the three IL models, in points A to E, i.e,
$d_z = \{13.8, 15.5, 18.0, 19.8, 25.8\}$~{\AA}. A common feature of all investigated IL models is the formation of {\it fixed} cationic layers along the whole length of the solid plates (top and bottom). The fixed layers and their stability are result of strong LJ interactions between the plates and ions. In general, the smallest particles form the first layer next to the plates: for TM3 these are tail particles (which are part of the cation-tail pair), while for TM5 and TM9 models these particles are the cations. The consecutive layers are formed inside the interplate gap via combined volume exclusion and Coulombic interactions,  and their ordering is
consistent with the fixed layers. As a result, tails migrate to the plates in TM3 model, they mix with the cationic layer when cation-tail
dimer is symmetric in TM5 model, and finally mix into the anionic layer when they are large in TM9 model. Since Coulombic interactions cause the layering with alternating charge sign, layers of anions always separate the cation layers.

We focus on analysing the changes in the segment between the points A and D, i.e., the interval \textit{I}. The normal force $F_z$ changes  rapidly and non-monotonously with $d_z$ in the interval \textit{I}, cf. Figure~\ref{fig:fd_static}. 
For the minimum of $F_z$ in the vicinity of point A, i.e., for plate-to-plate distance $d^{\rm A}_z = 13.8$~{\AA}, we can observe a well-defined anionic layer in Figure~\ref{fig:ionic_distribution} (corresponding snapshots of configurations are given in SI).
The most interesting change takes place during the transition A$\rightarrow$B when the single layer of anions is split into two layers, cf. Figures~\ref{fig:ionic_distribution}.
As a result, the normal force $F_z$ increases and reaches a local maximum in the proximity of point $B$, i.e., for plate-to-plate distance $d^{\rm B}_z = 15.5$~{\AA}. 
We observe that additional anion-cation pairs are pulled inside the gap in Figure~\ref{fig:ionic_density}. We also observe that the two anionic layers in Figure~\ref{fig:ionic_distribution} for point B and the one for point A have the same maximum number density. 
As we increase $d_z$ further, the number of anionic layers confined inside the gap remains unchanged and the normal load $F_z$ drops slowly. 
At the same time, the number of ions inside the gap steadily increases with the gap width. Nevertheless, this increase is not sufficient to keep the density of IL inside of the
gap constant (cf. Figure~\ref{fig:ionic_density}). 
Looking into changes in the spatial distribution
of IL components, as more cation-anion pairs
are pulled into the gap (going from A$\rightarrow$E), we observe a steady increase of the concentration of anions in the layer next to the bottom plate. In the case
of TM5 model we have an increase from $n_{\rm TM5}^{\rm A}=18$~atoms/nm$^3$ to $n_{\rm TM5}^{\rm D}=27$~atoms/nm$^3$, cf. Figure~\ref{fig:ionic_distribution}. 
When we further look at configuration snapshots for TM3 and TM5 model, a formation of additional layers inside the gap is visible, between the points C and D. This can also be clearly observed in Figure~\ref{fig:ionic_distribution} and results in smaller maximum around $d_z=$19~\AA, in Figure~\ref{fig:fd_static}. We can conclude that the form of the normal force-plate distance characteristic is not correlated with the number density of the IL molecules inside the gap, but the layer formation seen in Figure~\ref{fig:ionic_distribution}.

As the interplate distance $d_z$ increases further, from point
D to E, we notice additional cations in the middle of the gap
and formation of a third cationic layer in all three systems.  We can
make an interesting observation: for all three models the tails in the middle of the confinement are grouped in three regions: 
with cations at $z=$34~\AA, and in the middle between cationic 
and anionic layers, i.e., $z=$30,~38~\AA, cf. in Figure~\ref{fig:ionic_distribution}. This outcome is reminiscent of the findings from the Ref.~\cite{perkin2011self} where the authors have experimentally obtained the formation of the tail--to--tail bilayer of cationic dimers in case the alkyl chain length is
oversized.

%%%%%%%%%%%%%%%%%%%%%%
%FIG. 10
\begin{figure*}[ht!]
\includegraphics[width = 15 cm]{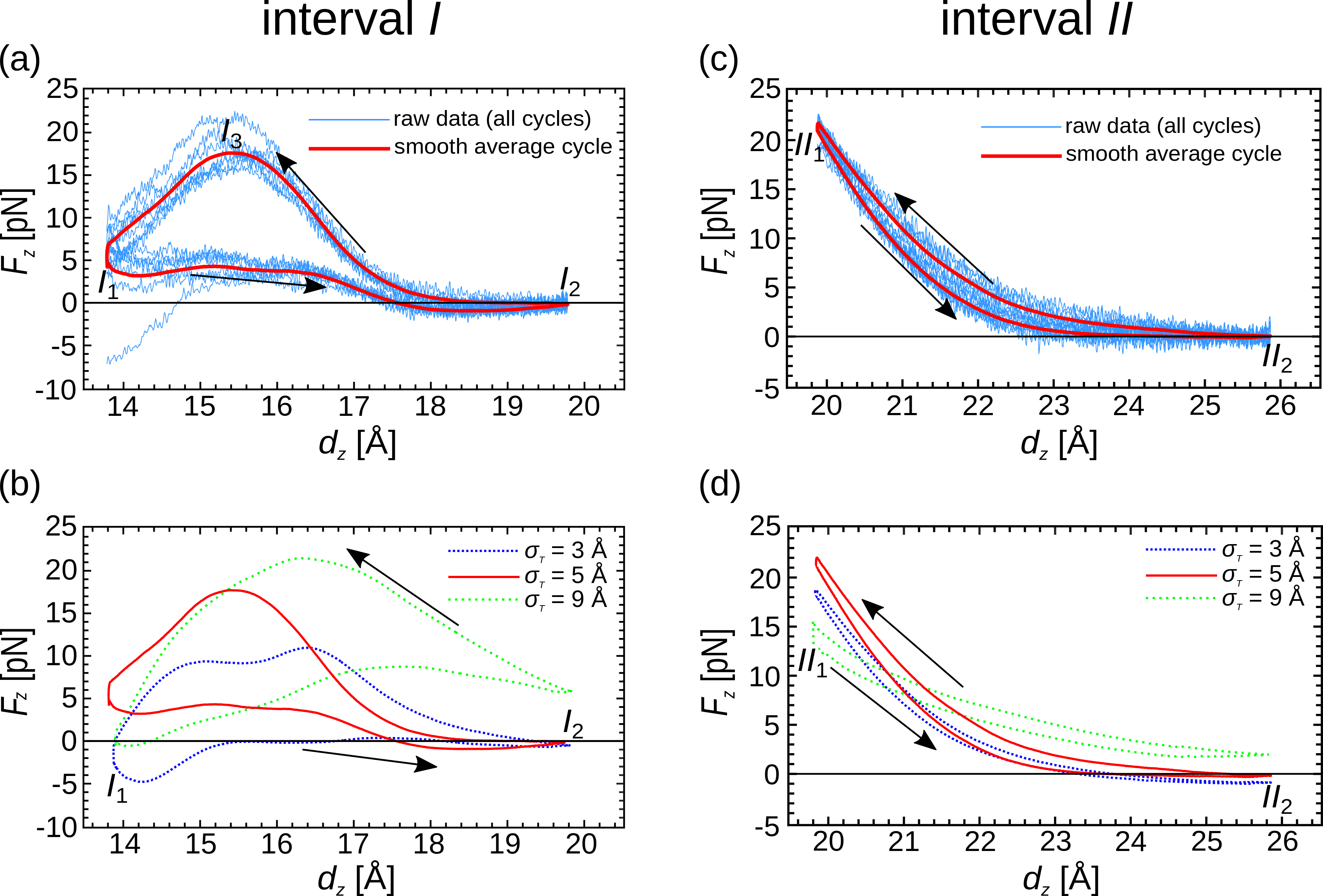}
\caption{The results of dynamic extension--compression cycles are shown for the intervals \textit{I} and \textit{II}.
In the panels (a) and (c) we present dynamic $F_z\left(d_z\right)$ characteristic in case of TM5 model, for the intervals \textit{I} and \textit{II}, respectively; thin lines represent the hystereses of ten dynamic cycles, 
solid line on top of them is the smooth average hysteresis. There is also a solid horizontal line which corresponds to $F_z = 0$.
In (a) points \textit{I}$_1$, \textit{I}$_2$, \textit{I}$_3$ denote representative points: \textit{I}$_1$ - starting point, \textit{I}$_2$ - ending point, \textit{I}$_3$ - global maximum of the $F_z\left(d_z\right)$ curve.   
In (c) points \textit{II}$_1$ and \textit{II}$_2$ denote representative points: \textit{II}$_1$ - starting point and \textit{II}$_2$ - ending point. 
The arrows show the direction of hysteresis (extension \textit{I/II}$_1$ $\rightarrow$ \textit{I/II}$_2$ followed by compression \textit{I/II}$_2$ $\rightarrow$ \textit{I/II}$_1$).  
In the panels (b) and (d) we show together smooth average hystereses $F_z\left(d_z\right)$ of our three TM~s, for the intervals \textit{I} and \textit{II}, respectively. 
Starting and ending points and arrows are denoted, analogous to the panels (a) and (c). 
}
\label{fig:dynamic_intervals}
\end{figure*}
%%%%%%%%%%%%%%%%%%%%%%

\subsubsection{IL crystallinity: influence of the gap}
We show the $xy$ cross-section snapshots in Figure~\ref{fig:confxy} in order to observe the IL's in-plane structure at the cross-section just below the top plate. We mark the boundaries of the top plate spatial region with the vertical dashed lines. The central area of the panels in the figure corresponds to the interplate gap region and it represents a half of the total cross-section's width in the $y$ direction, while the remaining area corresponds to the lateral reservoirs. 
The solid lines mark the orientation of crystal grains in those areas, where we can observe the presence of structural ordering.  
In the case of the TM3 model, we observe the presence of partial triangular ordering only at point B when the structure is the most compressed.
We do not notice any crystallization for symmetric dimers (TM5 model), which confirms that the symmetric tail prevents ordering both under confinement and in the bulk. Contrary to the previous two cases, we observe crystallization for all configurations with the large tail (TM9 model). Additionally, we observe changes in the type of crystalline structure. 
While in the lateral reservoirs a triangular lattice arrangement is always present, depending on the amount of compression we observe triangular lattice arrangements in points A and D and square lattice arrangements in points B and C. Even more surprisingly, the order is lost when the tail--to--tail bilayer is formed in point E.   

\subsection{Cyclic extension and compression of confined IL}
The top plate was moved between the two limiting points of the intervals \textit{I} ($d_z^{\rm A}\leq d_z \leq d_z^{\rm D}$) and \textit{II} ($d_z^{\rm D} \leq d_z \leq d_z^{\rm E}$).
We have investigated the dynamic behaviour of the confined IL thin film during the cyclic movement of the top plate along the $z$ axis, i.e., the interplate gap was periodically extended (\textit{extension} half--cycle)
and compressed (\textit{compression} half--cycle). We have investigated our system at three velocities $V_z = \left\{0.1, 1, 10\right\}$~m/s, but we did not observe any velocity dependent differences in the system behaviour.
The confined ionic liquid lubricant responds to the cyclic movement of the top plate with a hysteresis in normal force $F_z\left(d_z\right)$ shown in Figure~\ref{fig:dynamic_intervals}.
We present the detailed results of the TM5 model dynamic behaviour in (a) and (c) panels of Figure~\ref{fig:dynamic_intervals}.
Also, in (b) and (d) panels of the same figure, we present together smooth average cycles of our three IL models (TM3, TM5, and TM9).

\subsubsection{Narrow gap: normal force hysteresis}
We will now discuss in detail the response of the TM5 model to the cyclic motion of the top plate, in the interval \textit{I} shown in Figure~\ref{fig:dynamic_intervals}(a). 
Ten cycles of compression-extension are shown (thin lines) with an {\it average} cycle superimposed on them (thick line). 
We identify three points of interest: \{\textit{I}$_1$, \textit{I}$_2$, \textit{I}$_3$\}, i.e., the two terminal points of the cycle and the point with the maximal normal force, respectively. 
These three points also correspond to the points $\left\{A, D, B\right\}$ respectively, in the quasi-static characteristics shown in Figure~\ref{fig:fd_static}. Point \textit{I}$_3$ corresponds 
to the maximum of normal force $F_z$ both in the cyclic compression cycle and in the static characteristic of TM5 model which makes the comparison more straightforward.
 
The normal force $F_z$ decreases down to a value close to zero during the \textit{extension} half of the cycle \textit{I}$_1\rightarrow$\textit{I}$_2$. 
The anion-cation pairs are pulled into the gap from the lateral reservoirs as the gap is extended and at point~\textit{I}$_2$ an additional anionic layer is fully formed inside the gap. 
Actually, instead of the two fixed layers of cations which shared one anionic layer, we obtain two separate anionic layers. The total number of ions pulled in is about 60~atoms or 0.22~atoms/(nm$^2$ns) at 1~m/s plate linear speed. 
In the first part of the compression half-cycle, \textit{I}$_2\rightarrow$\textit{I}$_3$, the ions are compressed and the density and the normal force $F_z$ increase. Somewhat surprisingly, we observe that an equal number of ions flows out while the normal force increases, i.e., \textit{I}$_2\rightarrow$\textit{I}$_3$ and during its sharp drop \textit{I}$_3\rightarrow$\textit{I}$_1$ (cf. Figure~\ref{fig:ionic_density}). The sharp decrease of the normal force $F_z$ in the segment \textit{I}$_3\rightarrow$\textit{I}$_1$ is therefore a result of two processes: out-flow of the ions from the gap and the collapse of the anionic double layer and its rearrangement into a single anionic layer. The resulting final density $\rho^{\rm dyn}_{\rm IL}=$1.95~atoms/nm$^3$ of the system is slightly higher than in static case $\rho^{\rm stat}_{\rm IL}=$1.85~atoms/nm$^3$, cf. Figure~\ref{fig:ionic_density}. The value of the normal force $F_z$ at point \textit{I}$_1$ is similar, i.e., $F_z = 4$~pN in both static and dynamic case.

In Figure~\ref{fig:dynamic_intervals}(b), we observe that each one of the three investigated ionic liquids (TM3, TM5, and TM9) exhibits different behaviour in the average $F_z\left(d_z\right)$ cycle during the extension and compression half-cycle. First, at the onset of the extension half-cycle, i.e. in the point \textit{I}$_1$,
the normal force $F_z$ has a positive value for symmetric cations (TM5 model), it is close to zero for large tails (TM9 model), and it is negative for small tails (TM3 model).
Somewhat surprisingly, the normal force increases for both TM ILs with asymmetric cations (TM3/TM9 models) while it decreases for symmetric cation (TM5 model). The reason for this behaviour is 
the strong interaction of the fixed layers of ions adjacent to the plates with the plate particles.
This interaction drives as many ions inside the gap as possible, resulting in the non-intuitive behaviour of the normal force due to an interplay of density and intra-IL LJ interactions. 
During the compression half-cycle for all three ILs the maximal normal force sustained was about $50\%$ smaller than in the quasi-static case, i.e., for TM5 model the maximal force is $F^{\rm max}_z$=17~pN in the dynamic case
and $F^{\rm max}_z$=40~pN in the static case (see Figures~\ref{fig:dynamic_intervals}(b) and ~\ref{fig:fd_static}). This observation indicates that the top plate's motion prevents the IL to fill the gap. We can also conclude that the mechanical response is mainly due to a rearrangement of the fixed layer and that the mobility of the IL molecules is too low to significantly increase the normal force resisting to the compression.  
If we analyse the rate of mass transfer outside of the gap, we conclude that there is a substantial slip, which results in a lower normal force. Without slip at a velocity $V_z = 1$~m/s, the normal force calculated based on the bulk
viscosity coefficient would be roughly two orders of magnitude higher.

%%%%%%%%%%%%%%%%%%%%%%
%FIG. 13
\begin{figure}[t!]
\includegraphics[width = 7 cm]{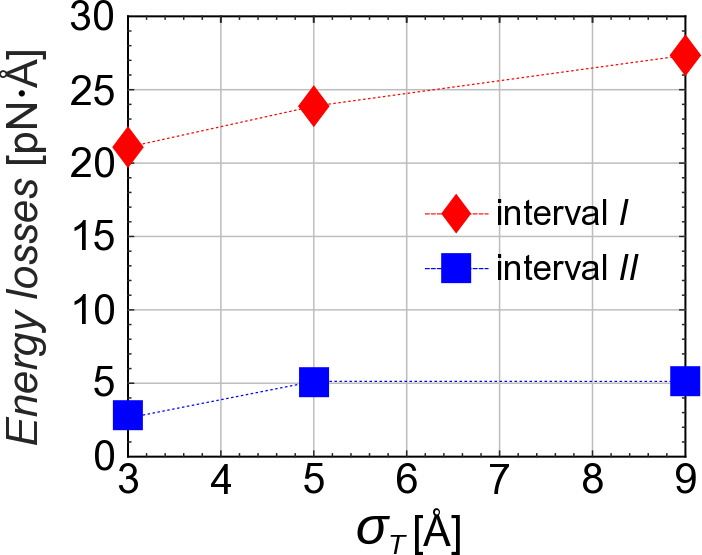}
\caption{Energy losses per average cycle in function of the tail size, for intervals \textit{I} and \textit{II} of dynamic extension--compression cycles.
}
\label{fig:energy_losses}
\end{figure}
%%%%%%%%%%%%%%%%%%%%%%

\subsubsection{Wide gap: monotonous force distance characteristics}
The expansion-compression force inter--plate distance characteristic for the interval~\textit{II} in case of TM5 model is given in Figure~\ref{fig:dynamic_intervals}(c). 
The difference from the quasi-static extension/compression in Figure~\ref{fig:fd_static} is the monotonous behaviour during the strike. 
The quasi-static characteristics in the interval \textit{II} featured local minima and maxima in the case of TM3 and TM5 models.
In the dynamic case, there are only two characteristic points (starting and ending point \{\textit{II}$_1$, \textit{II}$_2$\} and a monotonously changing normal force between them. 
In the {\it extension} half--cycle there is a continuous decrease of the normal force $F_z$ followed by its continuous increase in the {\it compression} half--cycle. 
The difference between the cycles in the normal force is small. In the dynamic characteristic of the interval~\textit{II} the layer structure is similar to the static case, i.e., two fixed layers stay-in-place
and the tail double layer is formed during the extension half-cycle (configuration snapshots are given in SI). In contrast to the interval~\textit{I}, the formation of the additional layer of tails is not a result of the ions flowing from the lateral reservoirs into the gap. The density inside the gap is 10\% higher in the dynamic case and a few atoms (less than 30) are displaced during the cycle. We should note that the gap is also 50\% larger in the interval~\textit{I} compared to the interval~\textit{II}, therefore the drop in density is even less striking. Actually, the cyclic motion has a tendency to increase the density inside the gap. Since there is no large displacement of the ions in and out of the gap in the interval~\textit{II}, there is also no maximum of the normal force $F_z$,
similar to the one we have seen in the case of the interval~\textit{I}, cf. Figure~\ref{fig:dynamic_intervals}(a).  
In order to make comparisons of different TM ionic liquid models, in Figure~\ref{fig:dynamic_intervals}(d)
we show together $F_z\left(d_z\right)$ average cycle dynamic characteristics of all three IL models (TM3, TM5, TM9) for the interval~\textit{II}. 
Compared to the interval~\textit{I}, the tail size does not have such pronounced impact on $F_z\left(d_z\right)$ hysteresis curves in the interval~\textit{II}.

\subsubsection{Energy losses due to cyclic expansion-compression}
At this point, we would like to quantify how do the processes arising during the dynamic cyclic movement of the top plate contribute to energy losses. 
We calculate the area covered during the extension-compression cycle (i.e., the area inside the $F_z\left(d_z\right)$ hysteresis).  
This area is equivalent to the work invested per average dynamic cycle, i.e., the hysteretic energy losses. 
We show the dependence of the energy losses on the tail size for both intervals \textit{I} and \textit{II} in Figure~\ref{fig:energy_losses}. 
We observe a clear tendency of the increase of the invested work per dynamic cycle, with the increase of the tail diameter. 
This is primarily due to the larger volume occupied by the tails resulting in larger normal forces resisting compression. There is a striking difference in the amount of invested work between the two intervals \textit{I} and \textit{II} (e.g. $27$~$\text{pN} \cdot \text{{\AA}}$ for the interval \textit{I} of TM9 model
compared to $5$~$\text{pN} \cdot \text{{\AA}}$ for the interval \textit{II} of TM9 model). 
This difference is proportional to the maximal normal force which is sustained by the systems in two the intervals (cf. Figure~\ref{fig:fd_static}).

%%%%%%%%%%%%%%%%%%%%%%
%FIG. 14
\begin{figure}[th!]
\includegraphics[width = 7 cm]{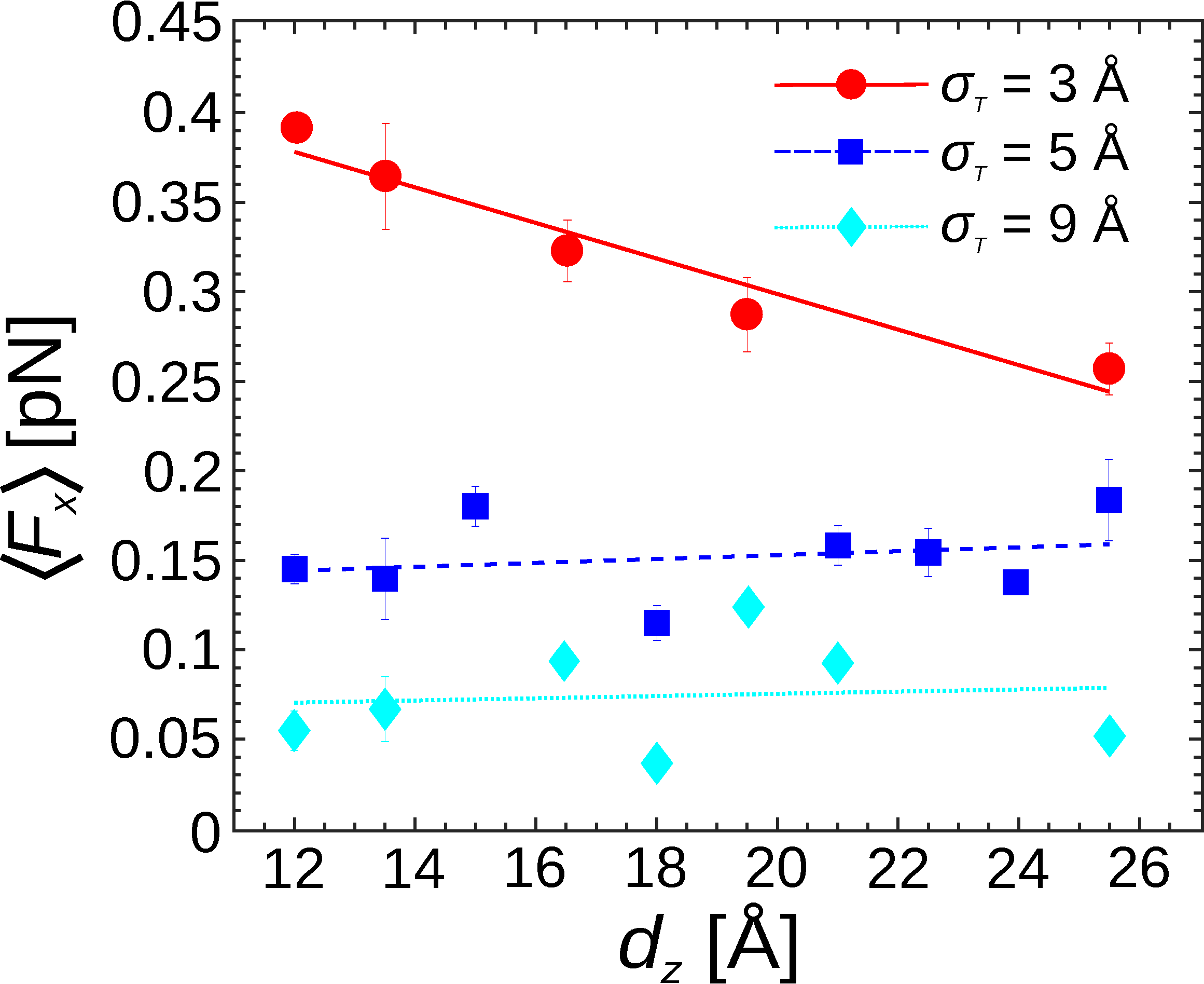}
\caption{Average frictional force $\langle F_x \rangle$ acting on the top plate as a function of the plate-to-plate distance $d_z$ for confined TM3, TM5 and TM9 ionic liquid lubricant. 
In case of TM3 model there is a clear linear dependence showing the decrease of frictional force intensity with the gap increase, while on the other side in case of TM5 and TM9 model
frictional force is practically constant and does not depend on the gap.
}
\label{fig:friction_gap}
\end{figure}
%%%%%%%%%%%%%%%%%%%%%%

%%%%%%%%%%%%%%%%%%%%%%
%FIG. 15
\begin{figure}[th!]
\includegraphics[width = 7 cm]{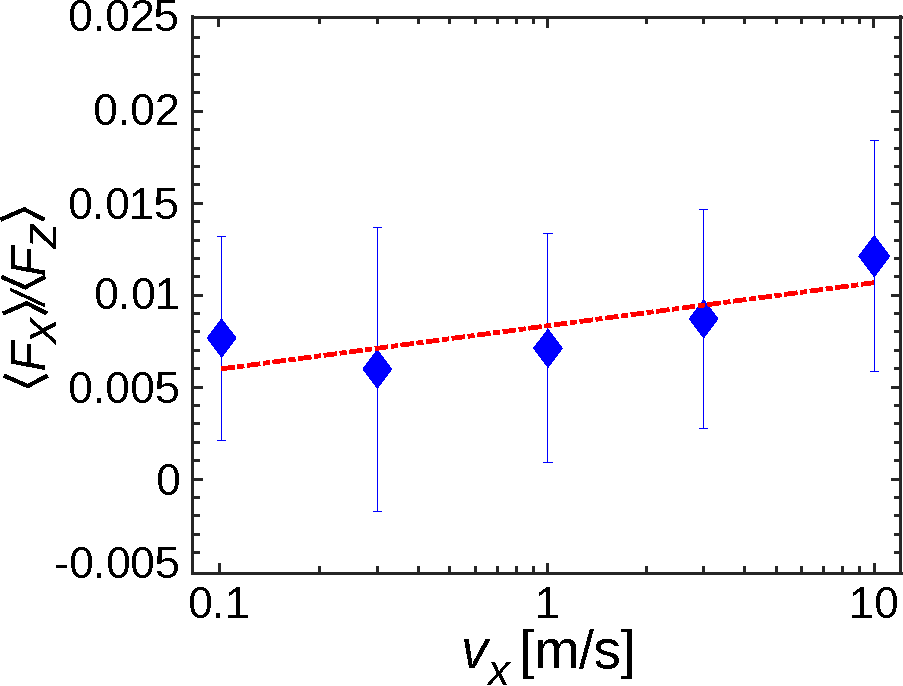}
\caption{Specific friction $\frac{\langle F_x \rangle}{\langle F_z \rangle}$ dependence on top plate's lateral velocity $V_x$ in case of TM5 model.
}
\label{fig:friction_velocity}
\end{figure}
%%%%%%%%%%%%%%%%%%%%%%
 
\subsection{Tribological Behaviour of Confined Ionic Liquid}
We have conducted static and dynamic characteristic analysis of the three generic IL models, focusing on the influence of their molecular structure on the anti-wear performance. 
In order to obtain a full picture, it is crucial to determine IL's friction behaviour under different shear conditions. 
In this section we apply a relative motion between the plates by moving the top plate along the $x$-axis (see Figure~\ref{fig:schemeTM}) and we observe the resulting frictional force (also along $x$-axis, i.e., $F_x$).
We have performed two types of friction simulations: $(i)$ at a constant top plate's velocity $V_x = 2$~m/s, the simulations are performed
at different fixed values of the gap: $d_z =$12~{\AA} to 25.5~{\AA} and $(ii)$ at a fixed gap $d_z = 15$~{\AA} top plate's lateral velocity takes five different values:
$V_x = \{0.1, 0.3, 1.0, 3.0, 10.0\}$~m/s. In all friction simulations, the total distance covered by the top plate was $\Delta_x = 100$~{\AA} in the $x$ direction.

The dependence of the time-averaged frictional force $\langle F_x \rangle$ on the interplate gap $d_z$ for the three IL models is shown in Figure~\ref{fig:friction_gap}. The points obtained in the simulations are shown as markers. Linear fits through these points are provided as visual guides. For the TM3 model, we observe a decrease of the frictional force $\langle F_x \rangle$ with the size of the gap. On the other hand, the frictional force weakly depends on the interplate gap width in case of TM5 and TM9 model ILs. Both the TM3 and TM9 have high zero shear-rate (Green-Kubo) bulk viscosities correlated with extent of their ordering, i.e., $\eta^{\rm GK}_{\rm TM3}>\eta^{\rm GK}_{\rm TM9}>\eta^{\rm GK}_{\rm TM5}$. When comparing with their tribological performance in a thin film we can conclude that there is no correlation since the TM5 IL has highest average frictional force. In Figure~\ref{fig:friction_velocity}, we show the dependence of specific friction $\langle F_x \rangle / \langle F_z \rangle$ on the top plate's lateral velocity $V_x$ in case of TM5 model. We obtain specific friction values of the order $\langle F_x \rangle / \langle F_z \rangle\approx0.01$ which are comparable to the result of \citeauthor{dold2013influence}~\cite{dold2013influence} for symmetric $\left[PF_6\right]^{-}$ anion. We observe also a similar tendency of decreasing friction force with respect to tail size, as reported in the same reference~\cite{dold2013influence}.

The specific friction $\langle F_x \rangle / \langle F_z \rangle$ is defined as the ratio of the time averaged frictional $\langle F_x \rangle$ and normal $\langle F_z \rangle$ force and it is different from the Coulombic friction coefficient $\mu=\partial{F_x} / \partial{F_z}$. Consistently with our previous results for model ionic liquids, we have observed a logarithmic dependence of specific friction
on the lateral velocity, cf. Ref.~\cite{tribint2017}. The numerical values are fitted to a linear function of the form $\langle F_x\rangle / \langle F_z\rangle = a \log(V_x/V_{\rm ref}) + b$, where $V_{\rm ref} = 1$~m/s. 
The coefficients of the linear fit took those values: $a = 0.001, b = 0.008$. A reasonable fit to the linear regression curve can be observed.
The logarithmic dependence indicates typical elastohydrodynamic lubrication (EHL) conditions~\cite{Bair2016}.

\vspace{0.6cm}
\section{Discussion}
Ionic liquids interact via long--ranged Coulombic forces and their models require high--performance computational resources. This opens a question of the minimal model needed to capture the properties of the molecular processes governing lubrication mechanisms and the macroscopic performance relevant for engineering applications. In this paper, we investigate a generic tailed-model (TM) of ionic liquids (ILs) which includes: an asymmetric cation consisting of a positively charged head and a neutral tail of variable size and a large spherical negatively charged anion. We observe that, though simple, this model results in striking differences of the equilibrium IL bulk structure governed by the tail size relative to cationic head: $(i)$ simple cubic lattice for the small tail, $(ii)$ liquid-like state for symmetric cation-tail dimer, and $(iii)$ molecular layer structure for the large tail. 

We have investigated the influence of the molecular structure of cation dimer on the response of three ILs to confinement and mechanical strain using molecular dynamics simulations. Properties of three IL models are compared in and out of equilibrium. We have related the evolution of normal force with inter-plate distance to the changes in the number and structure of the confined IL layers. We find that density inside the gap has a secondary effect on the evolution of the normal force. We observe that symmetric molecule offsets intra-IL adhesion due to the ordering of IL. As a result, the thin layer of symmetric IL molecules exhibits non-negative normal force independent of the gap width. In analogy to the experimental observations, a tail--to--tail bilayer is formed for wide gaps in all three investigated model ILs. 
A mutual feature of all investigated model ILs is the formation of fixed (stable) layers of cations along the solid plates. The fixed layer formation is a result of strong LJ interaction between the plates and ions. A consequence of the fixed layer stability is a steep rise of the normal force at small interplate gaps. The steep rise of the normal force is an effect useful for preventing solid-solid contact and accompanying wear. The tails attached to the cations in the fixed layer migrate with increasing tail size.  Small tails form the first layer next to the plates. For symmetric molecules the tails form a mixed layer with cations, while large tails form a mixed layer with anions. 

We have explored the dynamic behaviour of IL thin film under cyclic extension--compression movements of the top plate. Two intervals of the interplate distances are investigated: narrow gap interval, where the anionic layer is split into two, and a wide gap interval where tail--to--tail layer is formed. For the narrow gap interval, we observe a significant flow of ions during the cyclic motion of the top plate. A sharp decrease of the normal force at the final stage of compression is not only a consequence of the density change due to the flow, but also is a result of merging of two anionic layers that repel each-other by the electrostatic Coulomb forces into a single one. The mobility of ions in/out of the gap is driven by their interaction with plates, i.e., filling of the fixed layers. As a result, for the narrow gap, the number of ions that entered the gap is 50\% smaller in the dynamic case than in the static case. This results in a smaller density inside the moving narrow gap. The difference between dynamic and static cases for the wide gap was even more striking. the number of ions that entered the gap is 80\% smaller in the dynamic case than in the static case. Surprisingly in wide gap the the density is higher in dynamic case due to lack of mobility of ions. The invested work per average cycle increases with the tail size increase for all three IL models. As one could expect, the invested work is higher for the narrow gap where the number of confined ions/ionic layers changes during the cycle. Nevertheless, the low hysteretic losses suggest the presence of strong slip inside the gap facilitating in-- and out--flow of ions in the gap. An increase of the tail size reduces friction force in our model. Depending on the tail size, friction force decreases with increasing gap for small tails and it increases for large tails.
\vspace{0.6cm}
\section{Conclusion}

Understanding the interplay between the different processes taking place in thin lubricant films is important due to the conflicting demands imposed on how IL lubricant should behave in dynamic confinement. On the one hand, a high load-carrying capability requires strong adsorption of the lubricant to the surface, while on the other hand fast self-healing and low friction require high mobility/low viscosity. Our results confirm that behaviour of ILs in confinement can be unrelated to their bulk behaviour and therefore it should be possible to achieve simultaneously, typically conflicting, low friction and good anti--wear performance. A search for optimal IL lubricants, either using synthesis and test methods or state-of-the-art computer-aided molecular design methods~\cite{paduszyski2014viscosity}, should take into account the micro-scale properties of lubricating thin films (e.g., normal force vs. number of layers characteristics), in which the effects of molecular-level processes are more pronounced.
Directing the optimisation efforts to the micro-scale would enable us a better differentiation of the qualities of different ionic liquids.

\section{Acknowledgements}
M.D. and I.S. acknowledge the support of the Ministry of Education, Science and Technological Development of the Republic of Serbia under Project No. OI171017 and the support of COST Action MP1303. 
All computer simulations were performed on the PARADOX supercomputing facility at the Scientific Computing Laboratory of the Institute of Physics Belgrade, University of Belgrade, Serbia. 

\bibliography{references}

%%%%%%%%%%%%%%%%%%%%%%
%FIG. 6
\begin{figure}[ht!]
\includegraphics[width = 8 cm]{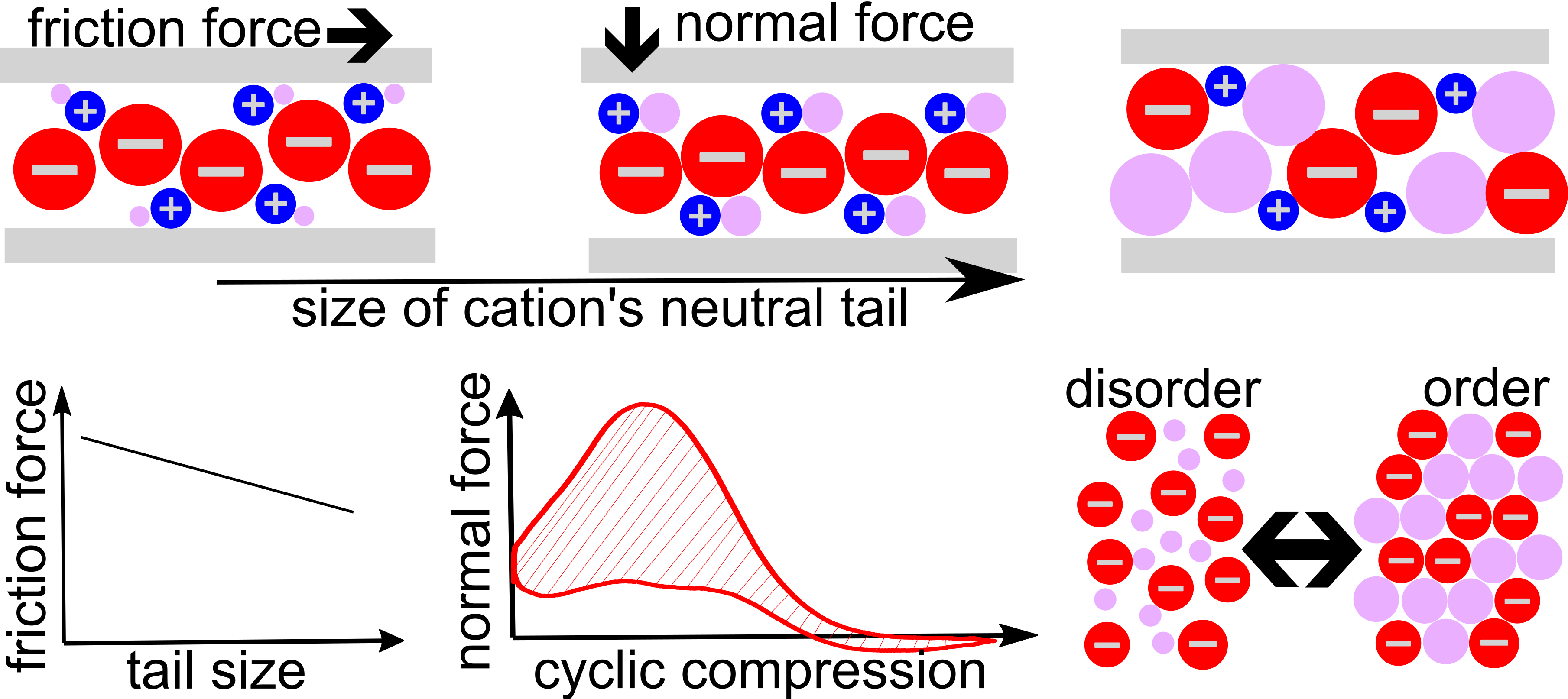}
\caption{Theoretical study of the influence of the geometry of the cation on the response of ionic liquid to confinement and mechanical strain is presented. The specific friction is low and friction force decreases with the tail size. The low hysteretic losses during the linear cyclic motion suggest strong slip inside the gap.}
\label{fig:ga}
\end{figure}
%%%%%%%%%%%%%%%%%%%%%%

\end{document}